\documentclass[twocolumn,useAMS]{mn2e}
\usepackage{amssymb} 
\input{epsf}
\def\etal{{\it et al.\ }}

\newcommand{\be}{\begin{equation}}
\newcommand{\ee}{\end{equation}}
\newcommand{\bea}{\begin{eqnarray}}
\newcommand{\eea}{\end{eqnarray}}

\title{The abundance of Lyman-\( \alpha  \) emitters in hierarchical models }

\author[M. Le Delliou {\it et al.}]
{M. Le Delliou
$^{1,2}$
\thanks{Morgan.LeDelliou@gamum2.in2p3.fr}
,  
  C. Lacey
$^{1,3}$
,
  C.M. Baugh
$^3$
, 
  B. Guiderdoni
$^4$%
,
  R. Bacon
$^1$
, \newauthor 
  H. Courtois
$^1$
,   
  T. Sousbie
$^1$
,
  S.L. Morris
$^3$
.\\
          $^1$Centre de Rech. Astron. de Lyon (CRAL), 
              9 avenue Charles Andr\'{e}, 69561 Saint Genis Laval Cedex, France. \\
	 $^2$GAM - UMR 5139 -  CNRS/in2p3,
	     CC 85,
	     U. M. II,
	     Place Eug\`ene Bataillon,
	     34095 Montpellier Cedex 5,
             France.\\
	  $^3$Institute for Computational Cosmology,
	      University of Durham, Science Laboratories.
	      South Road, Durham DH1 3LE, England.\\
	  $^4$Institut d'Astrophysique de Paris, CNRS,
	      98 bis Boulevard Arago,
	      75014 Paris,
	      France.\\
}
\pagerange{\pageref{firstpage}--\pageref{lastpage}}
\pubyear{2004}

\begin{document}

\maketitle

\label{firstpage}

\begin{abstract}
We present predictions for the abundance of Ly-$\alpha$ emitters in
hierarchical structure formation models. We use the {\tt GALFORM}
semi-analytical model of galaxy formation to explore the impact
on the predicted counts of varying assumptions about the escape
fraction of Ly-$\alpha$ photons, the redshift at which the universe
reionized and the cosmological density parameter. A model with a fixed
escape fraction gives a remarkably good match to the observed counts
over a wide redshift interval. The counts at bright fluxes are dominated 
by ongoing starbursts. We present predictions for the expected
counts in a typical observation with the Multi-Unit Spectroscopic
Explorer (MUSE) instrument proposed for the Very Large Telescope.
\end{abstract} 

\section{Introduction}
\label{sec:intro}
Dedicated narrow-band searches for Ly-$\alpha$ emitters have proven to
be very efficient at detecting high-redshift galaxies (e.g.  
Hu \& McMahon 
\etal 2003).  Objects found by this technique have to be confirmed
spectroscopically, to rule out possible low redshift interlopers that
may arise due to emission lines other than Ly-$\alpha$ falling within
the targetted wavelength interval.  Nevertheless, a significant
fraction of the detections appear to be bona-fide Ly-$\alpha$
emitters, and the number of objects accumulated to date by this
technique in the redshift interval $2.4<z<6.6$ is quite impressive.
The Ly-$\alpha$ emission line is also found in a significant fraction
of Lyman break galaxies (e.g. 
which are selected on the basis of their continuum emission.

The ubiquity of the Ly-$\alpha$ line is at face value surprising,
given that it is resonantly scattered by atomic hydrogen, and so is
easily absorbed by even a small amount of dust in a neutral gaseous
medium.  It is suspected that most Ly-$\alpha$ emitters have galactic
winds (as is the case with Lyman break galaxies) which allow
Ly-$\alpha$ photons to escape from the galaxy after only a limited
number of resonant scatterings (Kunth \etal 1998; Pettini \etal
2001). The Ly-$\alpha$ line typically shows an asymmetric profile
characteristic of such a process (e.g. Ahn 2004).  The physics of this
phenomenon is complicated, however, and remains poorly understood.

We present here the first predictions for the abundance of Ly-$\alpha$
emitters at different redshifts made using a model which follows the
formation and evolution of galaxies in a hierarchical universe. In
previous work, simple, ad-hoc prescriptions have been used to assign
star formation rates to dark matter haloes 
(Haiman \& Spaans 1999; Santos \etal 2004). 
In this Letter, we use a semi-analytical model
to make an {\em ab initio} calculation of the distribution of galaxy masses
and star formation rates at different redshifts (e.g. 
Cole \etal 1994; Kauffman \etal 1994; Baugh \etal 1998; Somerville \&
Primack 1999; Hatton \etal 2003).  The model we use is able to
reproduce the observed properties of galaxies both locally and at high
redshift \cite{galform,Baugh et al. 04}.  The abundance of Ly-$\alpha$
emitters is sensitive to the adopted cosmological model and to
astrophysical phenomena, such as the fraction of Ly-$\alpha$ photons
escaping from galaxies and the distribution of galactic dust.
Semi-analytical models are ideally suited to the exploration of such a
parameter space.

The semi-analytical model is described in Section \ref{assume}.  In
Section \ref{data}, we first present a compilation of the available
observational data on the abundance of Lyman-$\alpha$ emitters at
different redshifts, and then compare these data with the predictions
from our models. Finally, we present our conclusions in Section
\ref{concl}.

\begin{table}
\begin{tabular}{lllllllll}
\hline
\tiny{MODEL}&
\( \Omega  \)&
\( \Lambda  \)&
\( \sigma _{8} \)&
{\( z_{\rm reion.} \)}&
\( f_{\rm esc.} \)&
\tiny{MUSE COUNTS}\\
\hline
A&
\( 0.3 \)&
\( 0.7 \)&
\( 0.93 \)&
\( 10 \)&
\( 0.02 \)&
\(70\)\\
B&
\( 0.3 \)&
\( 0.7 \)&
\( 0.93 \)&
\( 10 \)&
\( 0.1 \)&
\(248\)\\
C&
\( 0.3 \)&
\( 0.7 \)&
\( 0.93 \)&
\( 10 \)&
{dust}&
\(366\)\\
D&
\( 0.3 \)&
\( 0.7 \)&
\( 0.93 \)&
\( 6 \)&
\( 0.02 \)&
\(163\)\\
E&
\( 0.3 \)&
\( 0.7 \)&
\( 0.93 \)&
\( 20 \)&
\( 0.02 \)&
\(77\)\\
F&
\( 0.2 \)&
\( 0.8 \)&
\( 1.15 \)&
\( 10 \)&
\( 0.02 \)&
\(145\)\\
G&
\( 1 \)&
\( 0 \)&
\( 0.52 \)&
\( 10 \)&
\( 0.02 \)&
\(59\)\\
\hline
\end{tabular}
\caption{The parameters of the semi-analytical models for which the
abundances of Ly-$\alpha$ emitters are predicted. The first column
gives the model label. The next three columns give the basic
cosmological parameters: the density parameter, $\Omega$, the
cosmological constant, $\Lambda$, and the amplitude of density
perturbations, as specified by $\sigma_{8}$ ($\sigma_{8}$ values are
taken from Eke \etal 1996).  In each case the baryon density is
$\Omega_{b}=0.04$ and the Hubble constant is $H_{0}=70 {\rm
kms}^{-1}{\rm Mpc}^{-1}$. Column five gives the redshift $z_{\rm
reion}$ at which the universe is assumed to reionize.  Column six
gives the fraction $f_{\rm esc}$ of Lyman-$\alpha$ photons that escape
from the model galaxies. In the case of model C, the escape fraction
is computed from a dust extinction model, as described in the
text. The final column gives the number counts of Lyman-$\alpha$
emitters for a reference {\tt MUSE} observation.  This is the number
of emitters in 1 square arcminute in the redshift interval $2.8<z<6.7$
brighter than $3.9\times10^{-19}{\rm erg s}^{-1}{\rm cm}^{-2}$.  }
\label{tab:Models}
\end{table}

\section{The model}
\label{assume}
We use the semi-analytical model of galaxy formation, {\tt GALFORM},
to make predictions for the abundance of Ly-$\alpha$ emitters as a
function of Ly-$\alpha$ flux and redshift. The {\tt GALFORM} model is
described in full in Cole \etal (2000) and Benson \etal (2003);
further details of the model used in this Letter are given in Baugh
\etal (2004).  The {\tt GALFORM} model computes the star formation
histories for the whole galaxy population.  The following steps are taken to
compute the Ly-$\alpha$ emission from a model galaxy: (i) The number
of Lyman continuum photons is computed from the star formation history
in the model galaxy and the stellar initial mass function (IMF). In
the Baugh \etal (2004) model, quiescent star formation in galactic
disks produces stars with a Kennicutt (1998) IMF, whereas bursts of
star formation triggered by galaxy mergers produce a flat
(``top-heavy'') IMF.  (ii) The luminosity of the Ly-$\alpha$ line is
computed assuming that all Lyman continuum photons are absorbed in HII
regions and produce Ly-$\alpha$ photons according to case B
recombination \cite{Osterbrock}.  (iii) The observed Ly-$\alpha$ line
emission depends on how many Ly-$\alpha$ photons escape from the
galaxy. We have taken two approaches to estimating the escape
fraction. In the first, we simply assume that a fixed fraction,
$f_{\rm esc}$, of Ly-$\alpha$ photons escape from the
galaxy. Physically, this might arise if a fraction of the Ly-$\alpha$
photons escape through holes in the galactic gas and dust
distribution.  In the second approach, we calculate the absorption of
Ly-$\alpha$ photons by a diffuse dust medium having the same spatial
distribution as the stars, ignoring resonant scattering by neutral
hydrogen. This may mimic what occurs if resonant scattering is
suppressed by a galactic wind. We make a self-consistent calculation
of the dust optical depth of the model galaxies, using the predicted
galaxy scale length, gas mass and metallicity. Full details of this
dust extinction model can be found in Cole \etal(2000). We plan a more
detailed calculation of the escape fraction of Ly-$\alpha$ photons,
including the effects of resonant scattering and gas outflows, in a
future paper.

\begin{figure}
{\epsfxsize=8.truecm
\epsfbox[0 30 508 548]{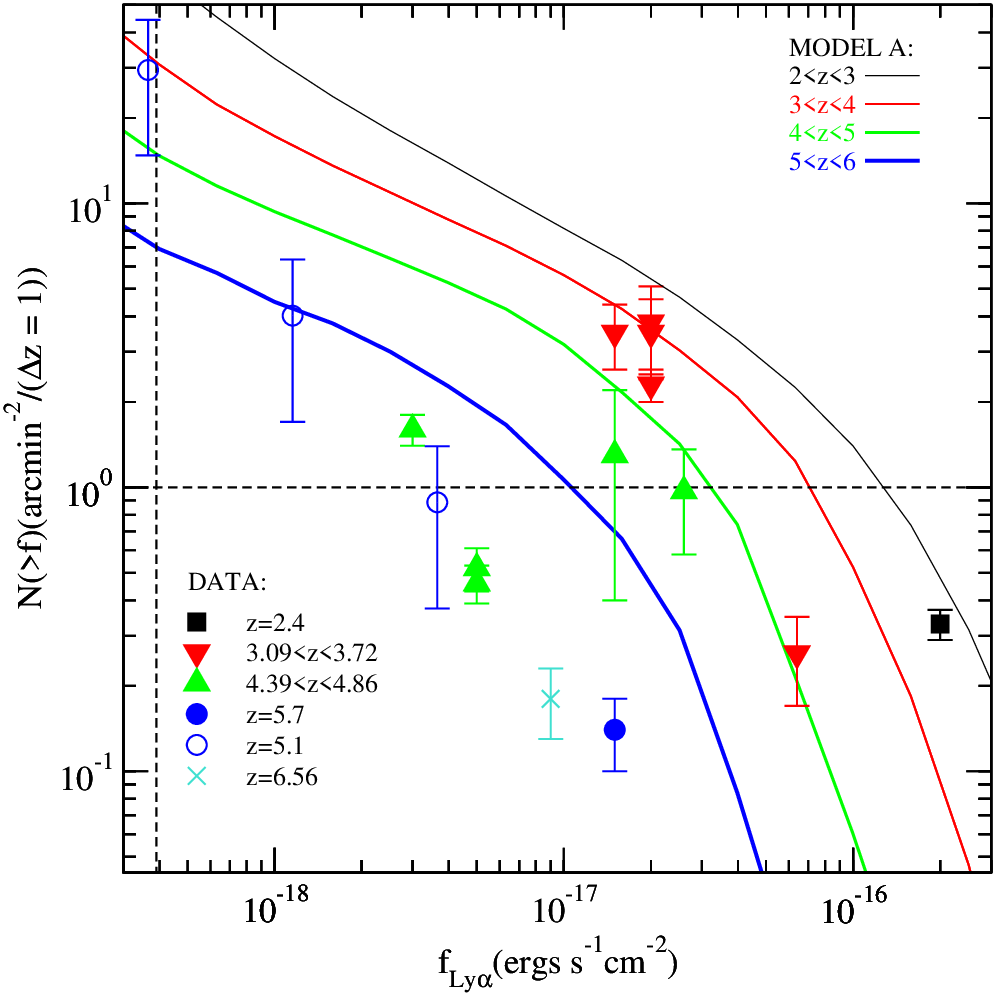}}
\caption{The predicted number of Ly-$\alpha$ emitters in model A,
  compared to the observational data compiled in Table \ref{tab:data}.
  The observational data are divided into different redshift ranges,
  as indicated by the lower key, and we have plotted only datapoints
  based on more than one galaxy. Model predictions are shown by lines,
  as indicated by the upper key. The thickness of the lines increases
  with redshift. The vertical and horizontal dashed lines indicate the
  sensitivity limits in flux and number density for a reference MUSE
  observation (see text).
\label{fig:data}
}
\end{figure}

\begin{figure}
{\epsfxsize=8.truecm
\epsfbox[0 30 529 550]{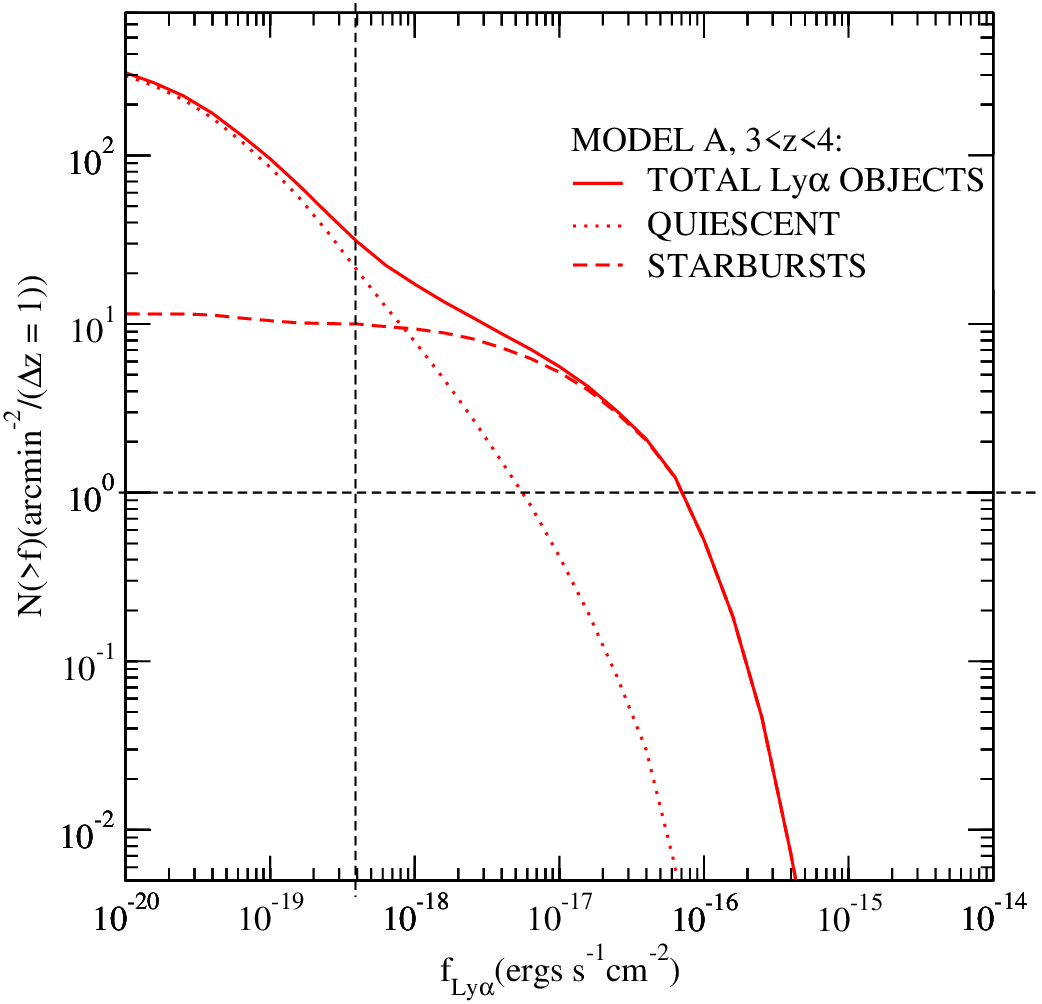}}
\caption{ The predicted number counts of Ly-$\alpha$ emitters in model A,
in the redshift range $z=3-4$. The contributions from quiescent star
forming galaxies and ongoing starbursts are plotted separately, as
indicated in the key. The vertical and horizontal dashed lines are
the same as in Fig.~\ref{fig:data}.  }
\label{fig:QB}
\end{figure}

We explore the impact on the abundances of Ly-$\alpha$ emitters of
varying the redshift at which the Universe was reionized, $z_{\rm
reion}$, which is still poorly constrained. 
\begin{table*}
\begin{tabular}
{llllllllllll}
\hline 
(1) & (2) & (3) & (4) & (5) & (6) & (7) & (8) & (9) & (10) & (11)\\
z&
\( \Delta z \)&
\( f \) &
\( \frac{d^{2}N}{dzd\Omega }(>f) \)&
\( \Delta \left( \frac{d^{2}N}{dzd\Omega }\right)  \) 
&
Nobj&
Area &
Fcorr &
method &
confirmation &
ref.\\
\hline
\( 2.42 \)&
\( 0.14 \)&
\( 20 \)&
\( 0.33 \)&
\( 0.04 \)&
\( 58 \)&
\( 1200 \)&
\( 0.65 \)&
NBF &
EW/colour &
Sti01\\
\hline 
\( 3.09 \)&
\( 0.07 \)&
\( 2 \)&
\( 2.3 \)$^*$&
\( 0.3 \)&
\( 12 \)$^*$&
\( 78 \)&
\( 0.94 \)&
NBF&
EW/colour &
Ste00\\
\( 3.13 \) &
\( 0.04 \)&
\( 2 \)&
\( 3.8 \) &
\( 1.3 \)&
\( 8 \)&
\( 49 \)&
\( 0.7 \)&
NBF &
spec on 10 &
K00\\
\( 3.43 \)&
\( 0.06 \)&
\( 1.5 \)&
\( 3.5 \)&
\( 0.9 \)&
\( 16 \)&
\( 75 \)&
\( 0.87 \)&
NBF &
spec on 15 &
H98\\
\( 3.72 \)&
\( 0.23 \)&
\( 6.4 \)&
\( 0.26 \)&
\( 0.09 \)&
\( 8 \)&
\( 130 \)&
\( 0.35 \)&
NBF &
colours &
F03\\
\hline
\( 4.39 \)&
\( 0.07 \)&
\( 2.6 \)&
\( 0.97 \)&
\( 0.11 \)&
\( 75 \)&
\( 1100 \)&
\( 0.33 \)&
NBF &
spec on 3 &
R00\\
\( 4.54 \)&
\( 0.06 \)&
\( 1.5 \)&
\( 1.3 \)&
\( 0.9 \)&
\( 2 \)&
24 &
\( 0.67 \)&
NBF &
spec on 3 &
H98\\
\( 4.79 \)&
\( 0.08 \)&
\( 0.5 \)&
\( 0.46 \)&
\( 0.07 \)&
\( 41 \)&
\( 1100 \)&
\( 0.8 \)&
NBF &
\( - \) &
S04\\
\( 4.86 \)&
\( 0.06 \)&
\( 0.5 \)&
\( 0.52 \)&
\( 0.09 \)&
\( 34 \)&
\( 1100 \)&
\( 0.8 \)&
NBF &
spec on 5 &
S03\\
\( 4.86 \)&
\( 0.06 \)&
\( 0.3 \)&
\( 1.6 \)&
\( 0.2 \)&
\( 52 \)&
\( 540 \)&
\( 0.6 \)&
NBF &
colours &
O03\\
\hline
\( 5.1 \)&
\( 1.0 \)&
\( 0.012 \)$^{**}$&
\( 48 \)$^{**}$&
\( 48 \)&
\( 1 \)&
\( 0.02 \)$^{**}$&
\( - \)  &
LS &
\( - \) &
Sa04\\
\( `` \)&
\( `` \)&
\( 0.037 \)$^{**}$&
\( 30 \)$^{**}$&
\( 15 \)&
\( 4 \)&
\( 0.14 \)$^{**}$&
\( - \) &
\( `` \)&
\( - \) &
``\\
\( `` \)&
\( `` \)&
\( 0.12 \)$^{**}$&
\( 4.0 \)$^{**}$&
\( 2.3 \)&
\( 3 \)&
\( 0.75 \)$^{**}$&
\( - \) &
\( `` \)&
\( - \) &
``\\
\( `` \)&
\( `` \)&
\( 0.37 \)$^{**}$&
\( 0.89 \)$^{**}$&
\( 0.51 \)&
\( 3 \)&
\( 3.4 \)$^{**}$&
\( - \) &
\( `` \)&
\( - \) &
``\\
\( `` \)&
\( `` \)&
\( 1.2 \)$^{**}$&
\( 0.14 \)$^{**}$&
\( 0.14 \)&
\( 1 \)&
\( 7.5 \)$^{**}$&
\( - \) &
\( `` \)&
\( - \) &
``\\
\( 5.3 \)&
\( 1.0 \)&
\( 2? \)&
\( 2.3 \)&
\( 1.0 \)&
\( 5 \)&
\( 2.2 \)&
\( - \)&
LS &
\( - \) &
D01\\
\( 5.7 \)&
\( 0.13 \)&
\( 1.5 \)&
\( 0.14 \)&
\( 0.04 \)&
\( 13 \)&
\( 710 \)&
\( 0.75 \)&
NBF &
spec on 4 &
R03\\
\hline
\( 6.56 \)&
\( 0.10 \)&
\( 0.6 \)$^{**}$&
\( 20 \)$^{**}$&
\( 20 \)&
\( 1 \)&
\( 0.46 \)$^{**}$&
\( 1 \)&
NBF &
spec on 1 &
H02\\
\( 6.56 \)&
\( 0.11 \)&
\( 0.9 \)&
\( 0.18 \)&
\( 0.05 \)&
\( 16 \)&
\( 810 \)&
\( 0.22 \)&
NBF &
spec on 9 &
K03\\
\hline
\end{tabular}

\caption{
\label{tab:data}
Data Compilation. The data are divided into unit redshift intervals:
the following symbols are used to denote data from each redshift
interval in the figures ($\blacksquare$ :[$z<3$], $\blacktriangledown$
:[$3<z<4$], $\blacktriangle$ :[$4<z<5$], {\LARGE $\circ$}:[$z=5.1$],
{\LARGE $\bullet$}:[$5<z<6$], {{\bf X}}:[$6<z<7$]).  {\bf Col.1:}
redshift; {\bf Col.2:} redshift interval; {\bf Col.3:} Ly-$\alpha$
flux (in $10^{-17} {\rm ergs\, cm}^{-2} {\rm s}^{-1}$); {\bf Col.4:}
cumulative counts per unit solid angle per unit redshift (in ${\rm
arcmin}^{-2}$); {\bf Col.5:} Poisson error on counts (in ${\rm
arcmin}^{-2}$); {\bf Col.6:} number of Ly-$\alpha$ emitters; {\bf
Col.7:} area of survey (in ${\rm arcmin}^{2}$); {\bf Col.8:} factor
applied to correct for contamination by low-z interlopers; {\bf
Col.9:} method (NBF=narrow band filter, LS=long-slit spectroscopy);
{\bf Col.10:} method used to reject or correct for low-z interlopers
(EW=equivalent width, spec on N = follow-up spectroscopy of N
objects); {\bf Col.11:} reference (D01: Dawson \etal 2001; F03: Fujita
\etal 2003; H98: Hu \etal 1998; H02: Hu \etal 2002; K03: Kodaira \etal
2003; K00: Kudritzki \etal 2000; O03: Ouchi \etal 2003; R00: Rhoads
\etal 2000; R03: Rhoads \etal 2003; Sa04: Santos \etal 2004; S03:
Shimasaku \etal 2003; S04: Shimasaku \etal 2004; Ste00: Steidel \etal
2000; Sti01: Stiavelli \etal 2001) }

\footnotesize
\noindent{$^*$ corrected for factor 6 overdensity
~~$^{**}$ corrected for gravitational lensing}
\end{table*}

We examine the consequences of making three choices: (i) $z_{\rm
reion}=10$, (ii) $z_{\rm reion}=20$ and (iii) $z_{\rm reion}=6$. The
first two values are consistent with the optical depth to last
scattering suggested by the WMAP measurement of the correlation
between microwave background temperature and polarization \cite{Kogut
et al. 03}. The latter value is suggested by the detection of a
Gunn-Peterson trough in the spectrum of a $z=6.28$ quasar by
Becker \etal (2001). In the model, gas is prevented from cooling
in haloes with circular velocities below $60{\rm kms}^{-1}$ for
redshifts $z<z_{\rm reion}$.  The models that we consider are listed
in Table \ref{tab:Models}. Models A-E reproduce the observed
luminosity function for the local galaxy population (e.g.  Baugh \etal
2004).  Finally, we also consider the effect on the counts of
Ly-$\alpha$ emitters of varying the cosmological density parameter,
$\Omega$ (models F and G).  Note that we have not attempted to vary
any other parameters of the {\tt GALFORM} model in these two cases to
force the model to reproduce the local galaxy luminosity function.

\section {The observational data and model predictions}
\label{data}

We list in Table \ref{tab:data} a compilation of published
observational data on number counts of Ly-$\alpha$ emitters at
different observed line fluxes and redshifts from ``blank field''
surveys that we will compare against our model predictions. Note that
some of the quantities listed in Table \ref{tab:data} are derived from
the information given in the original sources, so we present this
compilation in order to facilitate future comparisons between data and
model predictions by other authors.  We do not include data from
surveys targetted around high-z objects such as quasars (e.g. Fynbo
{\it et al.} 2001), since in this case the number density of
Ly-$\alpha$ emitters may be biased by an unknown factor. A few of the
surveys used spectroscopy to directly search for Ly-$\alpha$
emitters. However, most used narrow-band imaging to identify objects
having a strong emission line at the wavelength corresponding to
Ly-$\alpha$ at a particular redshift. Samples of objects obtained in
this way are generally contaminated by lower redshift galaxies for
which some other emission line (e.g. $[OII]3737$, $[OIII]5007$,
$H\alpha$) happens to fall at the same wavelength. This contamination
is typically estimated and removed using the equivalent width of the
emission line and/or broad-band colours, or from follow-up
spectroscopy on a subsample of the objects. The methods used in the
different surveys are indicated in the table.

\begin{figure*}
{\epsfxsize=13.truecm
\epsfbox[40 40 519 580]{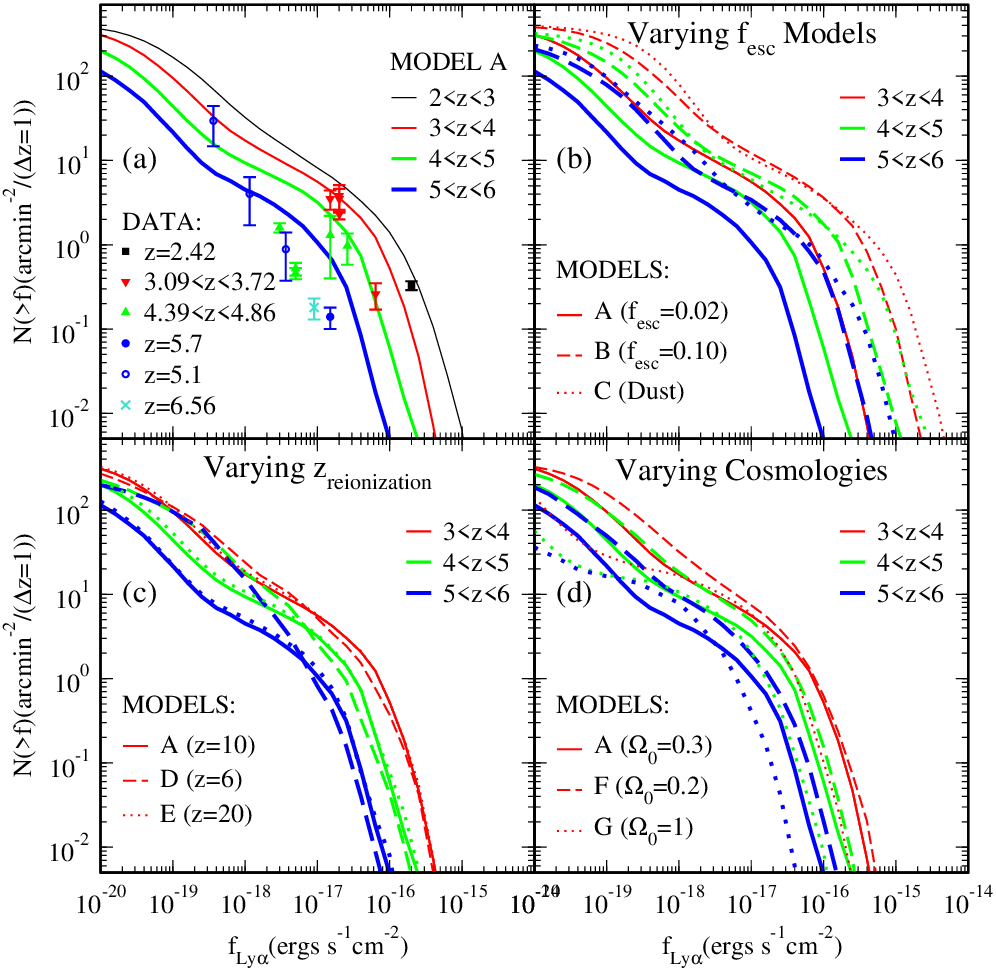}}
\caption{\label{fig:predictions} The predicted counts of Ly-$\alpha$
emitting galaxies, plotted in different redshift intervals, as
indicated by the key. (a) Shows the predictions for model A; a subset
of these are reproduced in each panel for reference.  The data from
Fig. 1 are also plotted here using the same symbols as before. The
remaining panels show the impact on the predictions of changing
different aspects of the model: (b) Varying the escape fraction or
using a dust model. (c) Varying the redshift of reionization. (d)
Varying the cosmological density parameter.  }
\end{figure*}

The observational data are plotted in Fig.~\ref{fig:data}, with
different symbols indicating data from a given unit redshift range.
The predictions of our fiducial model, A from Table~\ref{tab:Models},
are shown by the solid lines in Fig.~\ref{fig:data}. The escape
fraction $f_{\rm esc}$ was set to give a reasonable match to the
observed number counts at $z \sim 3$.  This simple model, in which the
escape fraction is independent of redshift and galaxy properties, does
a surprisingly good job of matching the observed counts at different
redshifts.  

The separate contributions to the counts of Ly-$\alpha$ emitters from
galaxies that are forming stars quiescently and from ongoing
starbursts (which in our model are triggered by galaxy mergers) are
shown in Fig.~\ref{fig:QB} for model A over the redshift interval
$z=3-4$.  Quiescently star forming galaxies dominate at fainter fluxes,
whereas starbursts account for the brighter Ly-$\alpha$ sources.  This
is largely due to the flat IMF assumed in starbursts, which typically
yields ten times the number of Lyman continuum photons for a given
amount of star formation, compared with the Kennicutt (1998) IMF that
we adopt for quiescent star formation.

In Fig.~\ref{fig:predictions}, we present the predicted counts of
Ly-$\alpha$ emitters as a function of redshift for the different {\tt
GALFORM} models listed in Table~\ref{tab:Models}. The results for
model A, our fiducial model, are reproduced for reference in each
panel. In Fig.~\ref{fig:predictions}(b) we show the effect of
increasing the escape fraction $f_{\rm esc}$ by a factor of
5. Fig.~\ref{fig:predictions}(b) also shows that assuming a fixed
escape fraction $f_{\rm esc}=0.1$ produces similar number counts to a
model in which the Ly-$\alpha$ photons are absorbed by diffuse dust
{\em without} resonant scattering, as might occur in a galactic wind.
Fig.~\ref{fig:predictions}(c) shows the impact on the predicted counts
of varying the redshift at which the universe is reionized.  For model
D, with $z_{\rm reion}=6$, the number of faint Ly-$\alpha$ sources
changes substantially for $5<z<6$ and $4<z<5$, compared with the cases
where $z_{\rm reion}\ge10$. This is because in model D gas is still
able to cool in low mass haloes (i.e. with circular velocities below
$60{\rm kms}^{-1}$) up to $z=6$, and is available to form stars and
thus generate Lyman continuum photons.  Finally,
Fig.~\ref{fig:predictions}(d) illustrates the effect of changing the
cosmological density parameter whilst retaining a flat universe. The
changes in the model predictions in this case primarily reflect the
change in the normalisation of density fluctuations, as specified by
$\sigma_{8}$ in Table~\ref{tab:Models}, which is adjusted to reproduce
the local abundance of rich clusters when the density parameter is
varied.

\section{Discussion and conclusions}\label{concl}
The conclusion of this first study to look at the predictions of
hierarchical models for the number of Ly-$\alpha$ emitters is that
simple models do remarkably well at reproducing the observed counts.
Assuming that typically just 2 \% of the Ly-$\alpha$ photons escape
from high-z galaxies, a value chosen to match the observed counts at
$z \sim 3$, is sufficient to give a reasonable match to the observed
counts at faint fluxes over the redshift interval $2<z<6$.

This study demonstrates the capability of semi-analytical modelling to
make predictions that can serve as an input into the design of new
instruments.  The Multi-Unit Spectroscopic Explorer ({\tt MUSE})
\cite{MUSE} has been proposed to ESO as a second-generation
instrument for the {\em Very Large Telescope}.  {\tt MUSE} will be
able to identify Ly-$\alpha$ emitters over a redshift interval
$2.8<z<6.7$ over a field of view of 1 arcmin$^2$.  An exposure of 80
hours will reach a 5 $\sigma$ sensitivity of $3.9 \times 10^{-19}$
erg/s/cm$^2$.  We predict that MUSE will be able to detect a large
number of such objects at this flux limit: around 70-400 per
arcmin$^2$ (see the final column of Table \ref{tab:Models}).
Observations with {\tt MUSE} will be able to exclude some of the
models we have considered, and therefore remove some of the
uncertainties in our modelling of Ly-\(\alpha \) emission. More
importantly, such observations will provide a critical test of our
ideas about star formation in objects at high redshifts.

\section*{Acknowledgments}
CGL and CMB thank their {\tt GALFORM} collaborators Andrew Benson,
Shaun Cole and Carlos Frenk for allowing them to use the model in this
paper. We acknowledge Bianca Poggianti for valuable discussions about
modelling emission lines.  We also thank the referee for providing a
speedy and helpful report that allowed us to improve the clarity of
this paper.  MLeD would like to thank the CRAL (Observatoire de Lyon)
for hospitality and financial support during the completion of this
work.  CMB and MLeD acknowledge support from the Royal Society. CGL is
funded by PPARC.

\end{document}